\documentclass{IEEEtran}
\usepackage{amsmath,amsfonts}
\usepackage{algorithmic}
\usepackage{algorithm}
\usepackage{array}
\usepackage[caption=false,font=normalsize,labelfont=sf,textfont=sf]{subfig}
\usepackage{hyperref,color,breqn,multirow}
\usepackage{textcomp,gensymb}
\usepackage{stfloats}
\usepackage{multirow}
\usepackage{url}
\usepackage{verbatim}
\usepackage{graphicx}
\usepackage{cite}
\usepackage{makecell}
\usepackage{booktabs}
\usepackage{amssymb}

\makeatletter
\renewcommand*{\eqref}[1]{\hyperref[{#1}]{\textup{(\ref*{#1})}}}
\newcommand*{\figref}[1]{\hyperref[{#1}]{\textup{Fig.~\ref*{#1}}}}
\newcommand*{\tabref}[1]{\hyperref[{#1}]{\textup{Table~\ref*{#1}}}}
\newcommand*{\secref}[1]{\hyperref[{#1}]{\textup{Section~\ref*{#1}}}}
\newcommand*{\mat}[1]{\overline{\overline#1}}
\makeatother

\begin{document}

\title{Reconfigurable Superdirective and Superabsorptive Aperiodic Metasurfaces}

\author{Yongming Li, Xikui Ma, Xuchen Wang, and Sergei~A.~Tretyakov,~\IEEEmembership{Fellow,~IEEE}


\thanks{Yongming Li is with the State Key Laboratory of Electrical Insulation and
Power Equipment, School of Electrical Engineering, 
Xi'an Jiaotong University, Xi'an 710049, China, and with the Department of Electronics
and Nanoengineering, Aalto University, P.O. Box 15500, FI-00076, Espoo, Finland (e-mail: yongmingli@stu.xjtu.edu.cn).}

\thanks{Xikui Ma is with the State Key Laboratory of Electrical Insulation and
Power Equipment, School of Electrical Engineering, 
Xi'an Jiaotong University, Xi'an 710049, China.}

\thanks{Xuchen Wang is with the College of Physics and Optoelectronic Engineeering, Harbin Engineering University, Harbin, 150001, China.}

\thanks{Sergei A. Tretyakov is with the Department of Electronics
and Nanoengineering, Aalto University, P.O. Box 15500, FI-00076, Espoo, Finland.}

}



\maketitle

\begin{abstract}

In this paper, we present a general theory of aperiodic subwavelength arrays for controlling electromagnetic waves. The considered platform is formed by an array of electrically small loaded scatterers above a ground plane. While the array is geometrically periodic, all the loads can be in general different, so that the distributions of currents induced by plane waves are not periodic. To allow analytical solutions, we study arrays of thin wires or strips loaded by bulk loads. We demonstrate a practical way of creating tunable and reconfigurable multifunctional devices, on examples of superdirective beam splitters, focusing lenses establishing subdiffraction focusing, and absorbers going beyond perfect absorption. Contrary to the constraints imposed by the Floquet theorem in periodic counterparts like periodic metasurfaces or metagratings, where a fixed angle of incidence and period dictate the propagating directions of reflected waves,  the proposed aperiodic designs allow controlling all propagating modes in any direction, which provides more freedom in manipulating electromagnetic waves. We hope that these results can be useful in multiple applications, such as telecommunications, radar techniques, signal processing, and energy harnessing.   
\end{abstract}


\section{Introduction}
\IEEEPARstart{I}{ntegrating} multiple and reconfigurable functions within a single electromagnetic device is of paramount importance, especially in wireless communications, due to the complexities and uncertainties of propagation environments. Recently, the reconfigurable intelligent surface (RIS) concept has drawn much attention due to the promise of creation of smart radio environments, based on manipulating reflection and transmission of electromagnetic waves. In this context, multiple reconfigurable functionalities, such as beamforming, focusing, and absorption, are highly desirable~\cite{elmossallamy2020reconfigurable, di2020smart,jiang2023simultaneously,meng2023efficient}.  Reflectarrays, a synthesis of conventional reflector antennas and phased arrays, are gaining popularity in modern electromagnetic and optical device designs due to their unique capabilities. Comprising a flat surface with periodic or aperiodic sub-wavelength structures, reflectarrays can modify and control the phase and amplitude of reflected electromagnetic waves. This level of control has opened doors to the realization of scanning reflectors capable of extraordinary feats. 

Over the past decade, in the constantly evolving field of electromagnetic wavefront manipulation, a lot of research has been conducted on metasurfaces and metagratings, particularly due to their periodic arrangements and their associated capabilities. These structures, with their distinctive periodic configurations, have displayed a plethora of sophisticated capabilities that have been imperative in numerous applications, such as anomalous reflection~\cite{asadchy2017flat, epstein2017unveiling, popov2018controlling, casolaro2019dynamic, liu2019intelligent, wang2020independent, popov2021non, ra2021metagratings, liu2023reflectarrays, kosulnikov2023molding,guo2018high,tan2023fully,tan2023efficient,tan2024fully}, focusing~\cite{yu2018design, rabinovich2019arbitrary, xu2021polarization, lou2022flat, wu2023multitarget, yang2023high,li2023terahertz},  power absorption~\cite{liu2019intelligent, li2019wideband, wang2020independent, lv2023wideband, tan2023design}, to name a few. In these works, to dynamically control the diffraction patterns, the incident directions or the array period have to be changed. For example, in Ref.~\cite{casolaro2019dynamic}, the authors present a theoretical formulation for complete manipulation of both reflected and transmitted fields using arrays of capacitively loaded strips. The proposed solution enables the design of electronically reconfigurable metagratings at microwave frequencies. In Ref.~\cite{liu2019intelligent}, the authors numerically show multiple reconfigurable functions, such as absorption and anomalous reflections by continuously tuning the local surface impedance. In Ref.~\cite{wang2020independent}, the authors  realize multiple functions by optimizing the scattering harmonics of arbitrary periodical space-modulated metasurface for multiple incident waves. However, for these periodic configurations, the diffraction orders of Floquet harmonics are determined by the angle of incidence and the period, e.g.~\cite{diaz2017generalized, diaz2021macroscopic}. This means that once the geometry of the structure is fixed, it is difficult to achieve continuous dynamic adjustment according to changes in the surrounding environment. 

Aperiodic reflectarrays and metasurfaces offer more general possibilities for control over reflected electromagnetic waves by manipulating the local properties of their constituting elements. 
In particular, they can realize other functionalities, such as focusing. 
Traditional aperiodic reflectarrays are designed using the phased-array principle and the locally periodic approximation (LPA), a method that has been in use since the 1960s \cite{Berry_reflectarray, Huang2008reflectarray}. In these devices, the adjustment aims to establish a prescribed phase variation of the induced current along the array plane by tuning the reactive load impedances or array elements. The optimization of these loads leverages the LPA, which numerically models the reflection coefficient from a single unit cell within an infinite periodic lattice under normal incidence. As with periodic metasurfaces crafted via LPA, the efficiency of reflectarray antennas diminishes significantly when the angles of incidence and reflection stray significantly from the conventional reflection law. 

Recently, non-local designs of reconfigurable aperiodic arrays for controlling the wavefront of reflected waves were considered in Refs.~\cite{popov2021non,li2024all,Sravan_arxiv}. In Ref.~\cite{popov2021non}, the authors introduced the concept of non-local reconfigurable sparse metasurface and experimentally validated  manipulation of electromagnetic waves in both near and far fields. However, such a design still suffers from low efficiency for large steering angles due to its sparse configuration (lack of engineering freedom), and the proposed platform cannot realize superdirective beamforming.
The results of studies of subwavelength-structured aperiodic arrays \cite{li2024all,Sravan_arxiv} demonstrated possibilities to realize nearly 100\% efficient
scanning anomalous reflectors. Moreover, it can be expected that by exploiting possibilities for evanescent fields optimization, it can become possible to create {superdirective } metasurface reflectors and absorbers that go beyond the fundamental limit of 100\% efficiency, reflecting and shaping more power than is incident on their surfaces. {We use the standard IEEE definition of superdirectivity as ``the condition that occurs when the antenna directivity is significantly higher than that
with the array or aperture uniformly excited'' \cite{standard}. Based on this, our figure of merit is the ratio of the directivity of the designed structure to the directivity of the corresponding uniformly excited surface of the same size and shape.  
The fact that superdirective and superabsorptive objects are possible is known for  a long time, e.g. \cite{Oseen,Schelkunoff,Haviland,bohren1983can}. Moreover, it is known that directivity of antennas can be in principle made arbitrarily high involving many resonant modes \cite{Harrington58}.} 
{The notion of superabsorption is closely related to that of superdirectivity. By definition, an electrically large object is superabsorbing if it absorbs more power than is incident on its body. In other words, the absorption cross section of a superabsorbing body is larger than the geometrical one. The maximal absorption cross section is related to the maximal directivity (e.g., \cite{Andersen,Liberal}), and, thus, it can be arbitrarily large.}
Known actual realizations are simple and practical only for electrically small resonant objects, e.g.~\cite{Haviland,ng2009metamaterial,valagiannopoulos2015electromagnetic,Shamonina,overcoming,Ziolk,wormhole}.

Realization of fully reconfigurable superdirective and/or superabsorbing panels remains a challenge. One of the first steps toward this goal was made in our initial study \cite{absorber_arxiv}, where we have demonstrated above 100\% absorption efficiency of subwavelength arrays for a given incident angle that can be adjusted by changing the array elements loads.

In this paper, we demonstrate reconfigurable flat arrays that exhibit superdirective or superabsorptive properties in several functionalties: beam splitters, absorbers, and focusing reflectors. We employ an analytical global optimization of bulk loads of subwavelength array elements. For conventional reflectarray, the distance between adjacent constituent elements is usually set to ${\lambda \over 2}$ to avoid grating lobes~\cite{Huang2008reflectarray}. However, such a sparse configuration, that is, with only one element in each ${\lambda \over 2}$-sized cell, cannot effectively control and optimize evanescent fields~\cite{li2024all}, which is needed for elimination parasitic scattering in anomalous and multi-channel reflectors~\cite{wang2020independent,zhong2023}, angular asymmetric absorption~\cite{wang2018extreme} and, especially, to realize superdirective properties.  In this work, dense strip arrays (the total number of strips $M$ in each ${\lambda \over 2}$-sized cell is larger than 1, i.e., $M>1$) are used to effectively control evanescent fields. Although evanescent modes do not carry power to the far zone, they are crucial in designing devices with high performance, e.g., anomalous reflectors~\cite{li2024all} and absorbers~\cite{wang2018extreme}. Here, 
the evanescent waves are controlled through numerical optimization, by setting appropriate objective functions for different functionalities. In order to illustrate the effectiveness of the developed non-local design method, several examples are presented. The numerical results demonstrate superdirective and superabsorptive performance for all considered devices. 
 
This paper is organized as follows. In \secref{sec:section2}, we define the problem and develop the theory. In \secref{sec:section3}, we introduce a method to design perfect beam splitters with an arbitrary power ratio between each angle with arbitrary directions. In \secref{sec:section4}, we present superabsorbing panels with angularly symmetric and asymmetric absorption properties. In \secref{sec:section5}, we show a {reflective} tunable focusing lens with a large numerical aperture and high focusing efficiency. \footnote{{The functionality of our reflective focusing lens is similar to focusing mirrors, for example, parabolic reflectors.}} Furthermore, we show how the method can be adapted to design focal lenses with multiple focal points and any desired intensity ratio. As an illustration, a {reflective} lens with two focal points is designed. Conclusions are drawn in the last section.

\section{Principle and methodology}
\label{sec:section2}

\begin{figure}[!htbp]
    \centering    \includegraphics{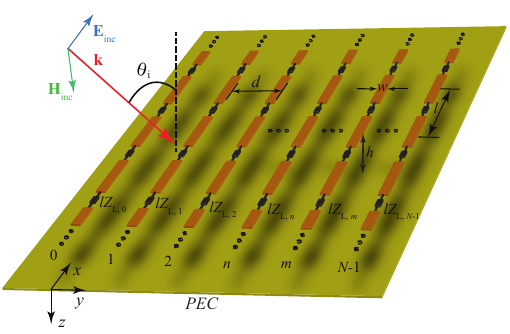}
    \caption{(a) An array of $N$ strips over an infinite ground plane under a plane-wave illumination at $\theta_{\rm i}$. The strips are loaded by bulk impedances inserted periodically with the period $l$. The width of the strips is $w$.  Both $l$ and $w$ are much smaller than the wavelength in free space. The periodically loaded strips can be modeled as homogeneous impedance strips with the impedance per unit length  $Z_{{\rm L}, n}$, where $n \in \{0, 1, 2, \dots, N-1\}$.}
    \label{fig:configuration}
\end{figure}

As a simple and analytically solvable platform, we consider a finite number of thin conducting strips placed along the $y$-axis and parallel to the $x$-axis at a distance $h$ above an infinite perfect electric conductor (PEC) ground plane. The ground plane is placed in the $z=0$ plane. The strips are equidistantly spaced and the distance between them (the geometric period) is $d$. The strips are periodically loaded by bulk reconfigurable loads, and it is assumed that the distance between the load insertions is much smaller than the wavelength of the incident wave. The array structure is depicted in \figref{fig:configuration}. The array is illuminated by a plane wave at the incidence angle $\theta_{\rm i}$. It is essential to note that the sign of the incident angle $\theta_{\rm i}$ is defined to be positive if the tangential component of the wavevector is along the positive direction of the $y$-axis and negative in the opposite case. The same convention is used for directions of reflected waves. The incident wave is assumed to be a TE-polarized plane wave with $\mathbf{E}_{\rm inc}(y,z) = E_0 e^{-j k_0( \sin \theta_{\rm i} y + \cos\theta_{\rm i} z)} \hat{x}$. 
In all numerical examples, the operation frequency is chosen as $10$~GHz, and the amplitude of the incident wave is set to $E_0=1$~V/m. The strip array above the ground plane $h={\lambda_0 \over 6}$, where $\lambda_0$ is the wavelength in free space. In the example studies of subwavelength arrays, the number of strips in each supercell of  ${\lambda \over 2}$ size is selected as $M=4$, and the total number of strips is 108. The width of the strips reads $\frac{\lambda_0}{100}$. In all of the examples below the first strip is placed at the position $y=0$, and the array width is $13.5\lambda_0$.  The time dependence is assumed to be in the form of $e^{j \omega t}$. 

Since the strips are very narrow, we model the current density flowing in strip $n$ as $\mathbf{J}_n (y,z) = I_n \delta(y-y_n,z + h) \hat{x}$, where $y_n = n d ~(n \in \{0, 1, 2, \dots, N-1\}$ is the position of the loaded strip and $\delta(x)$ is the Dirac delta function. According to Ohm's law, $I_n$ satisfies  
\begin{equation}
    Z_{{\rm L},n} I_n = E^{\rm ext}_x(y_n,-h) + \sum_{m=0,\\m \ne n}^{N-1} E_{nm} - Z_n I_n.
    \label{eq:Ohm_law}
\end{equation}
The left-hand side is the total electric field on the strip numbered $n$. The first term on the right-hand side is the external electric field, i.e., the superposition of the incident wave and the specular reflection of the incident wave from the ground plane, which can be defined as,
\begin{equation}
\mathbf{E}_{\rm ref} = - E_0 e^{-j k_0 (\sin \theta_{\rm i} y-\cos \theta_{\rm i} z)} \hat{x}
. \label{eq:reflected}
\end{equation}
Therefore, at the strip numbered $n$, $E_x^{\rm ext} (y_n,-h) =  j 2 E_0 \sin (k_0 \cos \theta_{\rm i} h) e^{- j k_0 \sin\theta_{\rm i} n d}$. The second term on the right-hand side is the field created by mutual interactions with the other strips, and it reads
\begin{equation}
    E_{nm} = -  Z_{nm} I_m,
\end{equation}
where $Z_{nm}$ ($m\ne n$) are mutual impedances between strip numbered $n$ and $m$ and are given by~\cite{li2024all}
\begin{equation}
    \small{
    Z_{nm} = \frac{k_0 \eta_0 }{4}  \left[ H_0^{(2)} \left(k_0 d_{nm} \right)  -  H_0^{(2)}\left(k_0 \sqrt{d_{nm}^2+ 4 h^2 } \right) \right]
    },
    \label{eq:mutual_impedance}
\end{equation}
where the distance between the strip numbered $n$ and $m$ is $d_{nm} = \left| y_m - y_n \right|$. The last term on the right-hand side is caused by its self-action, where $Z_n={k_0 \eta_0 \over 4} \left[H_0^{(2)}(k_0 r_{\rm eff}) - H_0^{(2)}(2 k_0 h) \right]$ are the self-impedances of the strips per unit length. In practice, thin conducting strips are modeled as equivalent round wires with the effective radius $r_{\rm eff} = w/4$~\cite{tretyakov2003analytical}.

For each loaded strip, the relationship between induced current and voltage has to satisfy \eqref{eq:Ohm_law}, which can be bridged through the load impedance matrix $\mat{Z}$ and given as
\begin{equation}
    \mat{Z} \cdot \vec{I} = \vec{U},
    \label{eq:matrix_Ohm}
\end{equation}
where the vector of induced currents is $\vec{I} = \left[ I_0, I_1, \dots, I_{N-1} \right]^{\rm T}$   and the voltage vector is $\vec{U} =  \left[ E^{\rm ext}_x(y_0,-h), E^{\rm ext}_x(y_1,-h), \dots, E^{\rm ext}_x(y_{N-1},-h) \right]^{\rm T}$. $\mat{Z} = \mat{Z}_{\rm s} + \mat{Z}_{\rm m} +\mat{Z}_{\rm L}$ is composed of the self-impedance matrix (a diagonal matrix) $\mat{Z}_{\rm s}={\rm diag}\left(Z_0, Z_1, \dots, Z_n, \dots, Z_{N-1} \right)$, the load impedance matrix $\mat{Z}_{\rm L} = {\rm diag}(Z_{{\rm L}, 0}, Z_{{\rm L}, 1},\dots, Z_{{\rm L}, N-1})$, and the mutual impedance matrix $\mat{Z}_{\rm m}$ with the matrix elements $Z_{nm}$ (see \eqref{eq:mutual_impedance}).

For a given set of load impedances $\vec{Z}_{\rm L} = (Z_{\rm L,0}, Z_{\rm L,1}, \dots, Z_{\rm L,N-1})$, the corresponding induced current distribution can be solved through \eqref{eq:matrix_Ohm}. Once the induced current distribution is known, the electric field generated by the strip array can be calculated as
\begin{align}
   \mathbf{E}^{\rm strips}=&- \frac{k_0 \eta_0 }{4}  \sum_{m=0}^{N-1} 
 I_m \left[ H_0^{(2)} \left(k_0 \sqrt{\left( y - y_m \right)^2+ \left( z+h \right)^2 } \right) \right.\notag\\
 &\left. -  H_0^{(2)}\left(k_0 \sqrt{\left( y - y_m \right)^2+ \left(z-h \right)^2 } \right) \right] \hat{x}. 
 \label{eq:electric_field_strips}
\end{align}
 The scattered electric field is calculated by subtracting the incident electric field from the total electric field.
 For our finite-sized array, the scattered electric field $\mathbf{E}^{\rm sca}$ is the sum of the field generated by the strips $\mathbf{E}^{\rm strips}$ and the specularly reflected field $\mathbf{E}_{\rm ref}$ \eqref{eq:reflected}, \emph{i.e.}, 
 \begin{equation}\mathbf{E}^{\rm sca} = \mathbf{E}^{\rm strips} + \mathbf{E}_{\rm ref}.
 \label{eq:scat}
 \end{equation}

\begin{figure}[!htbp]
    \centering
    \includegraphics{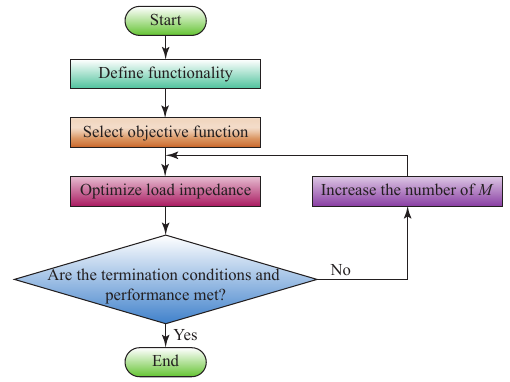}
    \caption{Block diagram of the design procedure for all exemplary devices proposed in this paper.}
    \label{fig:optimization}
\end{figure}

For the realization of different functionalities, we give the design workflow of the reflectarray-based design approach, which is depicted in \figref{fig:optimization}. The design process can be formulated as follows:

 Step 1: Determine the functionalities to be achieved, such as beam splitters, focusing lenses, or absorbers. For different functionalities, specify detailed requirements. For example, the number of angles, the power ratio between each angle, the number of focal points, focal length, the intensity ratio between each focal point, incident angles, and so forth.
 
 Step 2: Choose objective functions according to the different required functionality. For anomalous reflection, high directivity or high efficiency is highly desirable, while for absorber application, the requirement for reflection is the opposite. For a focusing lens, the power is expected to concentrate on the focal points, to improve the energy utilization efficiency.

 Step 3: Optimize load impedance distribution to realize the desired functionality until the termination conditions and performance are satisfied. Many global optimization methods have been used in metasurface-based gratings, such as the genetic algorithm (GA), differential evolution (DE), and particle swarm optimization (PSO)~ \cite{chipperfield1994genetic,popov2018controlling,campbell2019review, elsawy2020numerical}. Here, we use multi-population genetic algorithm (MPGA) to globally optimize connected loads. MPGA is an advanced method of GA, and it is more efficient than GA in searching complex, multimodal landscapes.

In general, augmenting the number of strips in each $\lambda/2$-spaced cell can enhance the performance of each required function, as it contributes additional flexibility for optimization. Using absorber designs as illustrative examples, we will delve into the impact of the variable $M$ on performance in~\secref{sec:section4}. 

To illustrate the effectiveness of our proposed aperiodic-based reflectarray design approach, three different functionalities with different examples are given. {In the numerical examples, we assume that realizations will be based on the use of varactor diodes, and set appropriate ranges of the load reactances. In particular, in the simulation, we restrict the reactive parts to negative values. The load reactance density range is [$-9.0 \times 10^5$,$- 1.0 \times 10^3$] $\Omega$/m. It is important to emphasize that the reactive impedance does not need to be purely capacitive; inductive reactance could also be included. Therefore, the reactance may be a combination of capacitive and inductive components, allowing greater flexibility and range in the impedance design.}

\section{Going beyond ``perfect'' beam splitters}
\label{sec:section3}
In our first examples, we demonstrate possibilities to create superdirective devices to control plane-wave reflections. 
Here, we present designs of beam splitters for reflections into any directions with arbitrary power ratios. The design method follows these steps:

\textit{Step 1:} Define the required number of reflected beams, the desired reflection directions, and the power division ratios. 

Considering a multiangle reflector, we denote the total number of reflection directions by $N_{\rm c}$ with the corresponding reflection angles  $\theta_{{\rm r}n}$. Next, we define the desired reflected power ratios between the reflected waves $n$ and the incident  for the corresponding infinite and periodic reflector illuminated by a plane wave:
\begin{equation}
    \rho_n = {P_n \over P_{\rm inc}}.
    \label{eq:power_ratio}
\end{equation}
Here, $P_n$ represents the power density (the amplitude of the normal component of the Poynting vector) of the wave reflected into direction  $n$, while $P_{\rm inc}$ represents the incident power density per unit area of the reflector. For an ideally performing infinite planar reflector, the power ratios satisfy  
\begin{equation}
     \sum_{n=1}^{N_{\rm c}} \rho_n = 1.
     \label{ref_multi}
 \end{equation}

\textit{Step 2}: Define the objective function in the optimization tool. 
The reference power ratios are defined above assuming that the array is a perfect infinite-sized multiangle reflector satisfying \eqref{ref_multi}. 

For an infinitely large perfect multiangle reflector, the induced current flowing on the reference reflector should contain several components, where one of them creates a plane wave that eliminates the specularly reflected wave and the others generate plane waves in the desired directions. Because these currents create plane waves, their amplitudes are the same for all strips. Let us denote the complex amplitude of the current that eliminates the specularly reflected wave as $I_\alpha$, and the others that generate the desired waves as $I_{{\beta}_1}, I_{{\beta}_2},\dots, I_{{\beta}_{N_{\rm c}}}$. In the phased array antenna theory, it is recognized that for the optimal anomalous reflection from non-superdirective arrays, the amplitudes of each component should have a uniform distribution and a linear varying phase along the whole array, e.g.~\cite{diaz2021macroscopic}. 
Hence, the reference induced currents on the strips satisfy  
\begin{equation}
    I_m = I_{\alpha} e^{- j k_0 \sin \theta_{\rm i} y_m} + \sum_{n=1}^{N_{\rm c}}I_{{\beta}_n} e^{- j k_0 \sin \theta_{{\rm r}_n} y_m},
    \label{eq:ind_currrents}
\end{equation}
where $m$ is the strip number.
The amplitude of the induced current $I_\alpha$ that eliminates specular reflection, reads (see derivations in Appendix A)
\begin{equation}
    I_\alpha = j \frac{E_0 d \cos \theta_{\rm i} }{\eta_0 \sin (k_0 \cos \theta_{\rm i} h)}.
    \label{eq:eliminate_spec}
\end{equation}
The amplitudes of the induced currents that generate the desired reflected waves $I_{\beta_n}$ equal (see derivations in Appendix A)
\begin{equation}
    \left| I_{\beta_n} \right| = \sqrt{\rho_n} \left| \frac{E_0 d \sqrt{\cos \theta_{\rm i} \cos \theta_{{\rm r}_n}}} {\eta_0 \sin (k_0 h \cos \theta_{{\rm r}_n})} \right|,  n \in \{1,~2,~\dots,~N_{\rm c}\},
    \label{eq:generate_des}
\end{equation}
where $\theta_{{\rm r}_n}$ is the desired reflection angle in angle $n$.  

It should be noted that to eliminate specular reflection, the phase of $I_\alpha$ varies along the array plane as that of the incident field (it is constant in case of normal illumination), while its initial phase is fixed by the phase of the incident wave. On the other hand, the initial phases of the currents corresponding to other reflection angles can be defined at will. In the presented example, the phases of the other reflection angles are selected as $0\degree$. 

After specifying the reference currents of a perfect infinite  reflector as in Eqs.~\eqref{eq:ind_currrents}--\eqref{eq:generate_des} (each component has a uniform amplitude and linearly varying phase), we can calculate the scattered fields from a finite-sized reflector assuming that on its surface the induced currents are the same as on the perfect infinite reflector. To do that, we use Eqs.~(\ref{eq:electric_field_strips}) and (\ref{eq:scat}). The far-zone electric field of the reference ``perfect'' reflector is defined as 
$E_{\rm refer}^{\rm far} (\theta_{{\rm r}_n})$, where $\theta_{{\rm r}_n}$ is the direction of angle $n$.

Next, we optimize reactive loads of finite-sized reflectors, comparing the achieved performance with that of the reference reflector defined above. 
To evaluate the achieved performance,  we introduce the power efficiency of angle $n$ in terms of the far field as 
\begin{equation}
    \zeta_n = \left| \frac{ E^{\rm far} (\theta_{{\rm r}_n}, \vec{Z}_{\rm L})} { E_{\rm refer}^{\rm far} (\theta_{{\rm r}_n})} \right| ^2,
    \label{eq:efficiency}
\end{equation}
where $E^{\rm far} (\theta_{{\rm r}_n},  \vec{Z}_{\rm L})$ is the far-zone field for a given set of load impedances $ \vec{Z}_{\rm L}$. The value of $E^{\rm far} (\theta_{{\rm r}_n},  \vec{Z}_{\rm L})$ is calculated by Eqs.~\ref{eq:matrix_Ohm}--\ref{eq:scat}, when the load impedance and incident conditions are given.
{In the literature, the efficiency of finite-size anomalous reflectors is usually defined with respect to the reflection from perfectly conducting plates of the same size and shape, with necessary corrections for the difference of the incidence and reflection angles, e.g. \cite{diaz2017generalized,Bipartite,Rabinovich}. The definition \eqref{eq:efficiency} has the same meaning, because in the physical optics approximation, the current induced at a perfectly conducting plate by a
plane wave is uniform in amplitude and has a linear phase dependence. This is the same as the current that creates the reference field in \eqref{eq:efficiency}. Since this reference current is not distorted close to the plate edges,  this is
the ideal case of the highest possible directivity of a non-superdirective reflector. We also note that in this definition there is no need to introduce corrections for the difference of the angles because the amplitude of the reference current is set to satisfy the energy conservation in reflections from infinite perfect anomalous reflectors.  }

Finally, the objective function is defined as
\begin{equation}
    O = \text{min.} \left\{ \left(\sum_{n=1}^{N_{\rm c}}  \left| \left| \frac{E^{\rm far}(\theta_{{\rm r}_n})}{ E^{\rm far}_{\rm refer}(\theta_{{\rm r}_n})} \right| - \gamma \right| \right) - \left( \gamma -1\right)  \right\}.
    \label{eq:reflector_obj2}
\end{equation}
Parameter $\gamma$ is one of the optimization variables, along with the load reactances of the array elements. The highest performance is reached for the maximized value of $\gamma>1$. The value $\gamma=1$ corresponds to the design goal of a conventional, not superdirective reflector whose performance is defined by \eqref{eq:power_ratio}, see~\cite{li2024all}. 

The objective function is specified as above with the aim of achieving the best attainable performance in terms of the highest possible far-field strength in the desired reflection directions, while the power ratios between the waves reflected into different directions are satisfied. The first term in~\eqref{eq:reflector_obj2} is intended to ensure the desired power division ratios, while the second term ensures maximization of the reflected fields in the desired directions. Both of these two terms have to be taken into account. If we only consider the first term in \eqref{eq:reflector_obj2}, the minimum value is zero, which occurs when the power ratio among each reflection angle is guaranteed, i.e., $\left|E^{\rm far}(\theta_{{\rm r}_n})\right| = \gamma \left|{ E^{\rm far}_{\rm refer}(\theta_{{\rm r}_n})} \right|$. However, this case does not necessarily be the optimal performance. In the process of optimization, the first term should be put in the first place, necessary weights can be put before the first term. While the first term obtains the minimum value, the second term will guarantee optimal performance. 

\textit{Step 3}: Optimize the load impedance distribution $\Vec{Z}_{\rm L}$ until the termination condition and performance is satisfied.

The objective function \eqref{eq:reflector_obj2} achieves its minimum value when $\gamma$ is maximized, meanwhile, the sum inside the absolute signs equals zero. After optimization, if $\gamma$ is greater than unity and the value of the objective function is smaller than $0$, the reflector exhibits superdirective properties, because the amplitudes of the reflected beams are higher than those for the perfect reference reflector defined by \eqref{eq:power_ratio}.
In this case, the efficiency  \eqref{eq:efficiency} is larger than 100\%.

Below we present examples of two-angle and three-angle beam splitters to illustrate the effectiveness of the proposed design method. Commercial software COMSOL Multiphysics is used to calculate the scattered electric field distribution and the power efficiencies. The configuration of the COMSOL Multiphysics simulations is shown in~Appendix~B.

\subsection{Two-angle beam splitters}
\

\subsubsection{Symmetric reflection angles}

\begin{figure}[!htbp]
    \centering
    \includegraphics{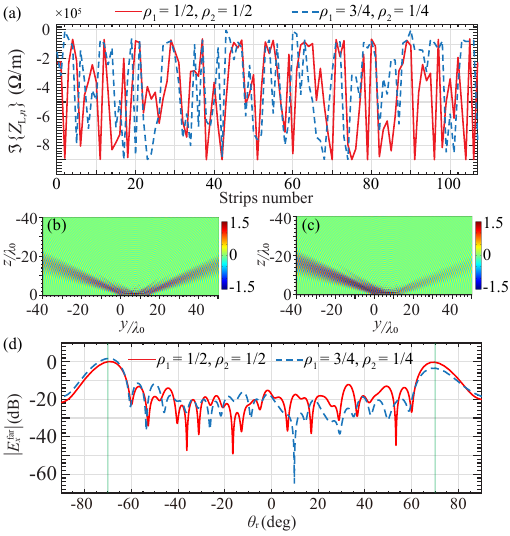}
    \caption{(a) Imaginary part of the optimized load impedance densities for symmetric two-angle beam splitter with equal power ratio represented by red curve and unequal power ratio represented by blue dashed curve. The real part of the scattered electric field distribution $\Re \{E_x^{\rm sca} \}~[{\rm V/m}]$ with (b) equal power ratio and (c) unequal power ratio. (d) The amplitude of the scattered electric field in the far zone as a function of reflection angle $\theta_{\rm r}$, where the red solid and blue dashed curves represent the equal power ratio, and unequal power ratio, respectively, while the green vertical line indicates the desired reflection directions of each angle.} 
    \label{fig:two_channels1}
\end{figure}

As the first example, we consider splitters that reflect normally incident waves into two mirror-symmetric beams with respect to the normal direction. In the considered examples, we set the reflection angles for Angle 1 and Angle 2 to $\theta_{{\rm r}_1} = -\theta_{{\rm r}_2} = -70 \degree$. 
Requiring equal distribution of power between the two reflected beams, that is,  $\rho_1 = 1/2,~\rho_2=1/2$. we calculate the reference induced current distribution as it is required in the denominator of the objective function \eqref{eq:reflector_obj2}. For the unit-amplitude incident electric field, the reference induced currents determined by \eqref{eq:eliminate_spec} and \eqref{eq:generate_des} equal $I_\alpha = j 1.5325\times 10^{-5}\text{A}$ and $\left| I_{\beta_1} \right|= \left| I_{\beta_2} \right| = 1.5656\times 10^{-5}\text{A}$. The reference fields ${ E_{\rm refer}^{\rm far} (\theta_{{\rm r}_n})}$ can be calculated  by~\eqref{eq:scat} in the far zone. With the obtained reference field, we use \eqref{eq:reflector_obj2} to optimize the distribution of load impedance reactances. The optimized load impedance distribution is shown in~\figref{fig:two_channels1}(a) by the red solid curve. The resulting scattered electric field pattern is shown in \figref{fig:two_channels1}(b).
To quantitatively evaluate the performance of the designed beam splitter, the amplitude of the scattered electric field in the far zone is depicted in \figref{fig:two_channels1}(d). It is shown that Angle 1 and Angle 2 have the same amplitude in the desired direction and have a symmetric distribution (see the solid red curve in \figref{fig:two_channels1}(d)). The efficiencies of both Angle 1 and Angle 2 are $107.6\%$, which corresponds to $\gamma=1.0373$. 

Next, we consider an example of a splitter with uneven ratio of the refected power, setting $\rho_1 = 3/4,~\rho_2=1/4$. The reference induced currents determined by \eqref{eq:eliminate_spec} and \eqref{eq:generate_des} in this case read  $I_\alpha = j 1.5325\times 10^{-5}\text{A}$, $\left| I_{\beta_1} \right|= 1.9175\times 10^{-5}\text{A}$, and $\left| I_{\beta_2} \right| = 1.1071 \times 10^{-5}\text{A}$. The optimized load impedance distribution is shown in~\figref{fig:two_channels1}(a) by the blue dashed curve. The resulting scattered electric field 
is shown in \figref{fig:two_channels1}(c), where the required stronger scattered electric fields can be noticed in the direction of Angle 1 as compared with Angle 2. The scattered electric field that radiates in the far zone is depicted in \figref{fig:two_channels1}(d) with a blue dashed curve. The efficiency of Angle 1 and Angle 2 are both $111.5\%$, which corresponds to $\gamma=1.0559$.

\subsubsection{Asymmetric angles}
For the case of asymmetric reflection angles, that is the two reflection angles do not have angular mirror symmetry along the broadside of the array. Here the reflection directions for Angle 1 and Angle 2 are chosen as $\theta_{{\rm r}_1} = -30 \degree$, $\theta_{{\rm r}_2} = 70 \degree$, respectively. 
\begin{figure}[!htbp]
    \centering
    \includegraphics{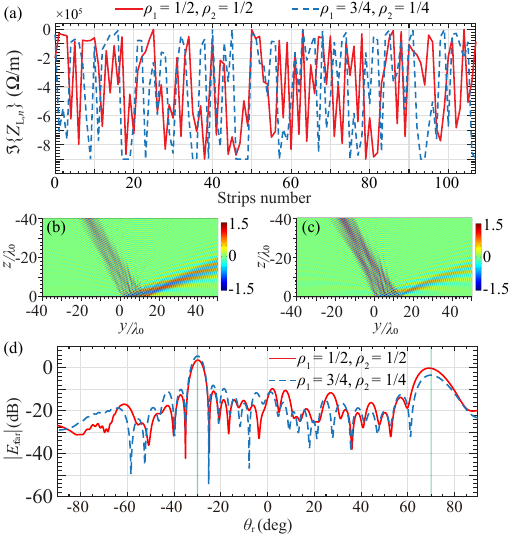}
    \caption{(a) Imaginary part of the optimized load impedance densities for the asymmetric two-angle beam splitter with equal power ratio represented by red curve and unequal power ratio represented by blue dashed curve. The real part of the scattered electric field distribution $\Re \{E_x^{\rm sca} \}~[{\rm V/m}]$ with (b) equal power ratio and (c) unequal power ratio. (d) The amplitude of the scattered electric field in the far zone as a function of reflection angle $\theta_{\rm r}$, where the red solid and blue dashed curves represent the equal power ratio, and unequal power ratio, respectively, while the green vertical line indicates the desired reflection directions of each angle.} 
    \label{fig:two_channels2}
\end{figure}

For the case of equal power ratio between the two reflection angles, i.e., $\rho_1 = 1/2,~\rho_2=1/2$, the referenced induced currents determined by \eqref{eq:eliminate_spec} and \eqref{eq:generate_des} are $I_\alpha = j 1.5325\times 10^{-5}\text{A}$, $\left| I_{\beta_1} \right|= 1.1089\times 10^{-5}\text{A}$, and $\left| I_{\beta_2} \right| = 1.5656\times 10^{-5}\text{A}$, respectively. The optimized load impedance densities are shown in~\figref{fig:two_channels2}(a) represented by the red solid curve. The resulting scattered electric fields are depicted in ~\figref{fig:two_channels2}(b). The scattered electric field that radiates in the far zone is depicted in \figref{fig:two_channels2}(d) represented by a blue dashed curve. The efficiency of Angle 1 and Angle 2 are both $105.8\%$, which corresponds to $\gamma=1.0287$.  

Regarding the case with unequal power ratio between the two reflection angles with $\rho_1 = 3/4,~\rho_2=1/4$, the referenced induced currents determined by \eqref{eq:eliminate_spec} and \eqref{eq:generate_des} are $I_\alpha = j 1.5325\times 10^{-5}\text{A}$, $\left| I_{\beta_1} \right|= 1.3581\times 10^{-5}\text{A}$, and $\left| I_{\beta_2} \right| = 1.1071\times 10^{-5}\text{A}$, respectively. The resulting scattered electric field is depicted in ~\figref{fig:two_channels2}(c). The far-field pattern is depicted in \figref{fig:two_channels2}(d) by a blue dashed curve. The efficiencies of Angle 1 and Angle 2 are both $105.1\%$, which corresponds to $\gamma=1.0252$.

\subsection{Three-angle beam splitters}
\begin{figure}[!htbp]
    \centering
    \includegraphics{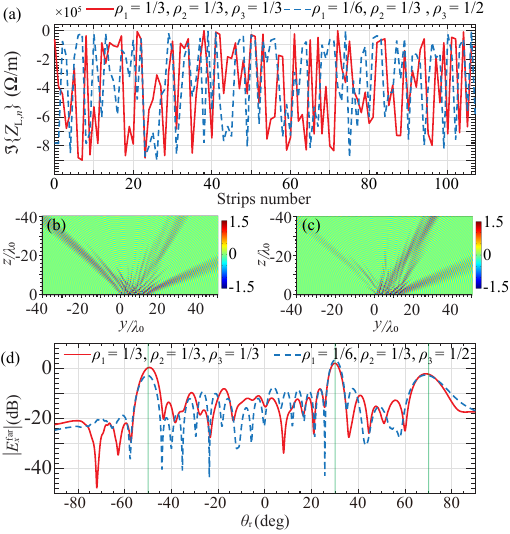}
    \caption{(a) Imaginary part of the optimized load impedance densities for the three-angle beam splitters. Red curve: uniform power division; blue dashed curve: uneven split. The real part of the scattered electric field distribution $\Re \{E_x^{\rm sca} \}~[{\rm V/m}]$ with (b) uniform split and (c) nonuniform split. (d) The amplitude of the scattered electric field in the far zone as a function of the observation angle $\theta_{\rm r}$, where the red solid and blue dashed curves represent the cases of the equal power ratios, and unequal power ratios, respectively, while the green vertical line indicates the desired reflection directions of each angle.}
    \label{fig:three_channels}
\end{figure}

Here, we consider an example design of a splitter that reflects normally incident waves into three directions: Angle 1 at $\theta_{{\rm r}_1} = -50\degree$, Angle 2 at $\theta_{{\rm r}_2} = 70\degree$, and Angle 3 at $\theta_{{\rm r}_3} = 30\degree$. 
For the case of the uniform power ratio distribution among different reflection directions, i.e., 
$\rho_1=1/3,~\rho_2=1/3,~ \rho_3=1/3$, the reference induced currents determined by \eqref{eq:eliminate_spec} and \eqref{eq:generate_des} are $I_\alpha = j 1.1494\times 10^{-5}\text{A}$, $\left| I_{\beta_1} \right|= 7.3906\times 10^{-6}\text{A}$, $\left| I_{\beta_2} \right| = 9.5875 \times 10^{-6}\text{A}$, and $\left| I_{\beta_3} \right| = 6.7904 \times 10^{-6}\text{A}$. The optimized load impedance densities are shown in~\figref{fig:three_channels}(a) represented by the red solid curve. The resulting scattered electric fields are depicted in ~\figref{fig:three_channels}(b). The far-field pattern is depicted in \figref{fig:three_channels}(d) as a red solid curve. The efficiencies of all three angles are equal to $103.1\%$, which corresponds to $\gamma=1.0155$. 

For the case of a different power ratio distribution among different angles, in the considered example of $\rho_1=1/6,~ \rho_2=1/3,~ \rho_3=1/2$, the referenced induced currents determined by \eqref{eq:eliminate_spec} and \eqref{eq:generate_des} equal $I_\alpha = j 1.1494\times 10^{-5}\text{A}$, $\left| I_{\beta_1} \right|= 5.2259\times 10^{-6}\text{A}$, $\left| I_{\beta_2} \right| = 9.5875 \times 10^{-6}\text{A}$, and $\left| I_{\beta_3} \right| = 8.3165 \times 10^{-6}\text{A}$. The optimized load impedance densities are shown in~\figref{fig:three_channels}(a) represented by the blue dashed curve. The resulting scattered electric fields are depicted in ~\figref{fig:three_channels}(c). The far-field pattern  is depicted in \figref{fig:three_channels}(d) by a blue dashed curve. The efficiencies of all three angles equal $100.3\%$, which corresponds to $\gamma=1.002$. It is important to note that the efficiency is related to the minimum value of the desired reflection angles. The efficiency will be higher for larger values of the smallest desired reflection angle.  This phenomenon has been noticed in our previous work~\cite{li2024all}. {The presented designs show superior performance, where the efficiency shows perfect scanning capabilities and superdirective
properties. Periodic metasurfaces and metagratings can exhibit theoretically perfect performance
at extreme angles even if the arrays are sparse~\cite{tan2023efficient,tan2023fully,tan2024fully}. However, they cannot scan, and for extreme
angle beam manipulation of scanning arrays, achieving high reflection efficiency requires effective generation of evanescent modes, which necessitates a tighter arrangement of elements. To the best of our knowledge, there are no superdirective designs by conventional
aperiodic metasurfaces~\cite{popov2021non,budhu2020perfectly}. }

{Regarding the implementation in practical applications, a similar geometry of strips arrays loaded by tunable impedances was recently experimentally realized in~\cite{popov2021non}. Moreover, actual realizations can be made based on other geometries, including conventional patch arrays, which have been practically realized by many teams. 
}

\section{Superdirective absorbers}
\label{sec:section4}

As another example of controlling plane waves with superdirective efficiencies, we consider planar finite-sized energy-absorbing plates. Such devices can have profound impact on various applications, such as energy harvesting, stealth, sensors, and so forth.  To realize absorption, the elements of the same array are connected to lossy loads, in contrast to purely reactive capacitive loads in the above examples of reflectarrays. Our goal here is to show that it is possible to realize superdirective absorption in an extremely wide range of the incident angles as well as angularly asymmetric absorbers, i.e., devices that combine the functionalities of retroreflectors and absorbers.  

Before delving into the design procedure, it is essential to define an appropriate parameter that measures absorption efficiency. As a reference scenario, we consider a uniform plane-wave illumination of an absorbing plate at an incident angle $\theta_{\rm i}$. The incident power is calculated as the power incident  on the geometric cross section of the strip array (per unit length along the strips), expressed as
$P_{\rm inc}={|E_0|^2 \over 2\eta_0}N d \cos \theta_{\rm i}$. The actual power delivered to the loads of the array elements (per unit length) reads $P_{\rm dis} = \frac{1}{2}\sum_{m=0}^{N-1} |I_m|^2 \Re\{ Z_{{\rm L},m} \}$. The absorptance, denoted as $A$, we define as the ratio between the delivered and dissipated power and the incident power:
\begin{equation}
    A = \frac{P_{\rm dis}}{P_{\rm inc}}.
    \label{eq:absorption}
\end{equation}
With this definition, for an infinite array, the achievable absorptance is always smaller than unity. However, here, for a finite-sized structure, the absorptance defined in \eqref{eq:absorption} can be greater than unity, i.e., $P_{\rm dis}>P_{\rm inc}$.  This phenomenon, known as superabsorption~\cite{bohren1983can,tribelsky,ng2009metamaterial, tretyakov2014maximizing,valagiannopoulos2015electromagnetic,overcoming,wormhole}, implies that the structure can receive more power than what is incident on its geometric cross-section. That is, the actually accepted power can exceed the power calculated by simply multiplying the incident power density with the cross-section area of the absorber.

\subsection{Angularly symmetric wide-angle absorption}

The design goal for absorbing panels is to extract as much power from the incident wave and deliver it to the resistive loads of the array elements. For a plane wave coming from the direction $\theta_{\rm i}$, the objective function is defined as
\begin{equation}
    O (\theta_{\rm i}) = \min. \left \{  \frac{P_{\rm inc}(\theta_{\rm i})} {P_{\rm dis}(\theta_{\rm i})} -1  \right \}.
    \label{eq:single_channel}
\end{equation}
The reason for this definition of an objective function is that the dissipated power can be greater than the power incident on its surface. When the value of the objective function is smaller than zero, it means superabsorption; for values larger than zero, lack of perfect absorption.

For angularly symmetric absorbers, that is, for structures that have the same absorption for incident angles $\pm \theta_{\rm i}$, the load impedance has a symmetric \textit{(even)} distribution with respect to the center of the array. This property reduces the number of variables to optimize by approximately half. If the total number of strips  $N$ is an even number, the total number of variables to optimize is also $N$. However, if $N$ is an odd number, then the total variables to optimize will be $N+1$. This calculation is due to the fact that each strip load impedance has both real and imaginary parts to be optimized.

Superabsorptive designs optimized for one angle of incidence $\pm \theta_{\rm i}$ were presented and discussed in our earlier work \cite{absorber_arxiv}. Here, we introduce and study wide-angle designs. For infinite absorbing boundaries, a ``perfect" absorber should fully absorb plane waves at all angles of incidence, which is the definition of an ideal black body. However, due to inherent spatial dispersion of any material interface, conventional periodic metasurface absorbers can fully absorb waves only for one specific incident angle. Away from this specific angle, the absorption rate decreases. Recent research has revealed that engineering spatial dispersion of a periodic metasurface can lead to perfect absorption for two \cite{8003278} or a wide range of incident angles \cite{wang2020independent}. However, these known solutions for infinite periodic boundaries are fundamentally limited by 100\%  absorptivity. Here, we demonstrate that within the general framework of finite-size aperiodic gratings, it is possible to design superabsorbing panels for multiple incident angles and even for a broad continuous angular spectrum.
 
\begin{figure}[!htbp]
    \centering
    \includegraphics{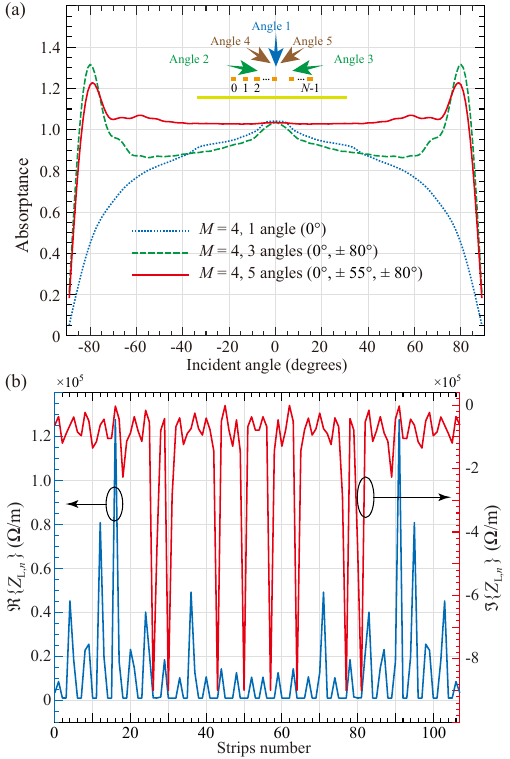}
    \caption{(a) Absorptance of symmetrical absorbers as a function of the incident angle. Angle 1 with the normal incidence, i.e., $\theta_{{\rm i}_1} = 0\degree$, while the Angle 2, 3 with $\theta_{{\rm i}_{2,3}} = \pm 80\degree$, and the Angle 4, 5 with $\theta_{{\rm i}_{4,5} }= \pm 55\degree$, respectively. The blue dotted line corresponds to the case with normal incidence. The green dashed line represents 3 angles absorber, which correspond to $\theta_{{\rm i}_1} = 0\degree,~\pm80\degree$. The red solid line represents 5 angles absorber, which correspond to $\theta_{{\rm i}_1} = 0\degree,~\pm55\degree,~\pm80\degree$.. (b) Real and imaginary parts of load impedance of each strip for the 5 angles ($0\degree$, $\pm 55\degree$, and $\pm 80\degree$) absorber. The blue and red curves, corresponding to the real and imaginary parts respectively, are displayed on the left and right axes. 
    }
    \label{fig:symmetric_absorption}
\end{figure}

 To obtain wide-angle absorption, multiple incident angles should be taken into account. In this case, the objective function is defined as 
\begin{equation}
    O (\theta_{\rm i_1}, \theta_{\rm i_2}, \dots, \theta_{{\rm i}_{N_{\rm c}}}) = \min. \left \{\sum_{k=1}^{N_{\rm c}} \left(  \frac{P_{\rm inc}(\theta_{{\rm i}_k})} {P_{\rm dis}(\theta_{{\rm i}_k})} -1 \right)  \right \},
\end{equation}
where $N_{\rm c}$ is the total number of incident plane waves that are taken into account. Optimization for only one incident angle \cite{absorber_arxiv} leads to superabsorption at the design angle, but with a narrow range of angles with such effective absorption. This is illustrated by an example design for normally incident waves, see the blue dotted line in \figref{fig:symmetric_absorption}(a). To widen the angular range of perfect absorption, we optimize for the normal incidence and one more angle, that of $80\degree$. In this case, the achieved absorptance as a function of the incident angle is depicted in \figref{fig:symmetric_absorption}(a) with the green dashed line. However, the absorptance at the normal direction becomes smaller while a much higher absorptance is achieved at $80\degree$. Moreover, a dip occurs in the directions of $\pm 55\degree$. To further improve the absorption valley, another incident angle of $\theta_{\rm i}=55\degree$ is also taken into consideration. As a result, the absorption curve becomes much wider and more uniform, as is shown by the red solid line in \figref{fig:symmetric_absorption}(a). 
Figure~\ref{fig:symmetric_absorption}(b) shows the corresponding distribution of the optimized load impedances. We observe fast variations of the load values over the plate surface, which is typical for superdirective devices.

\begin{figure}[ht]
    \centering
    \includegraphics{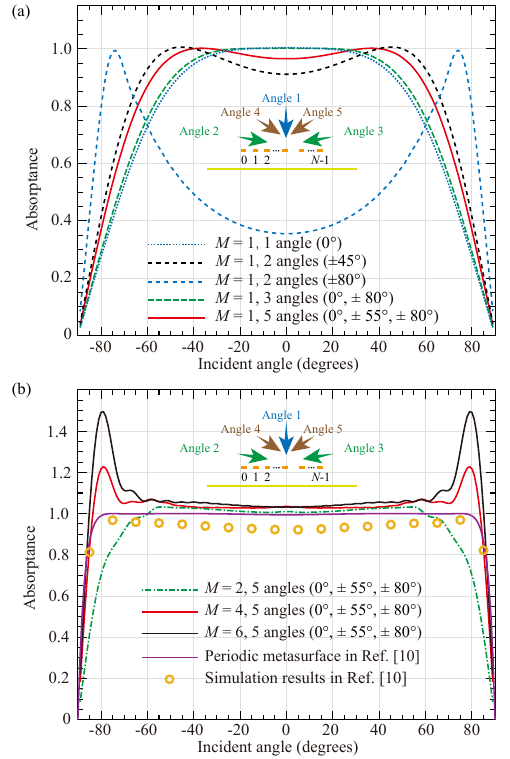}
    \caption{Absorptance as a function of the incident angle for the case (a) $M=1$ and (b) $M=2,~4,~6$. Also, a comparison with the previous work realized by periodic metasurfaces.}
    \label{fig:symmetric_absorber_comparison}
\end{figure}

Most importantly, in the whole wide range of the incident angles, approximately between $\theta_{\rm i}=\pm 85\degree$, the absorptance is larger than 100\%. 
To compare the performance with conventional realizations when the distance between the adjacent elements is $d=\lambda_0/2$, we do the same global optimization of a finite-size array for $M=1$. For this case of $\lambda/2$ distance between array elements, the absorptance as a function of incident angle is depicted in~\figref{fig:symmetric_absorber_comparison}(a). Superdirective absorption appears to be not realizable in this case. 
To realize superdirective absorbers, the array period must be smaller than  $\lambda_0/2$, to allow optimization of distribution of reactive fields.  Figure~\ref{fig:symmetric_absorber_comparison}(b) presents the results of a study of the achievable performance for different geometrical periods. We clearly see that the performance of absorbers can be significantly improved as we increase  $M$. We also compare our results with the previous work in Ref.~\cite{wang2020independent}, where absorption was realized by periodic metasurfaces. It can be seen that the designs introduced here achieve better performance, overcoming the fundamental limit of 100\% absorptance (see red and black curves in~\figref{fig:symmetric_absorber_comparison}(b)). In addition, the design method in Ref.~\cite{wang2020independent} suffers from absorptance decrease upon discretization of the optimized continuous surface impedance distribution (see orange circles in~\figref{fig:symmetric_absorber_comparison}(b)).

\subsection{Angularly asymmetric absorption}
In addition to realizations of wide-angle superabsorption, the proposed method can be extended to designs of angularly asymmetric absorbers. This is possible because the angular spectrum of absorption can be shaped by engineering evanescent waves~\cite{wang2018extreme}, which is also the key mechanism of superabsorption discussed above. Angularly asymmetric structures exhibit a broken symmetry of the absorption response for incident waves coming from the opposite directions with respect to the normal. That is, in this case the absorptivities for the incident angles $\pm \theta_{\rm i}$ are different and can be separately engineered. The effect of asymmetric absorption was found only in infinite periodic arrays~\cite{wang2018extreme}. Here, we will show that 
the use of finite-size aperiodic arrays with subwavelength separations between the elements offers full flexibility in engineering the asymmetry angle and absorption ratio, which cannot be realized by the conventional periodic metasurfaces or metagratings. Moreover, we find that also these devices exhibit superabsorption effect. 

The proposed structure exhibits a broken symmetry of the absorption response with different incident angles, that is, for some angle of incidence absorption is high, while for the opposite incident angle, a high-level retroreflection is observed. As an absorption asymmetry measure we define the retroreflection coefficient as 
\begin{equation}
    R = \left| \frac{E^{\rm far} (\theta_{\rm r})}{E^{\rm far}_{\rm refer}(\theta_{\rm r})} \right|,
    \label{eq:retroreflection_coeff}
\end{equation}
where $E^{\rm far}_{\rm refer}(\theta_{\rm r})$ is the reference far-zone electric field along the direction $\theta_{\rm r}$ created by a theoretically perfect reflector that carries the ideal current distribution of $I_\alpha$ and $I_\beta$ (with a uniform amplitude and linearly varying phase), defined  in~\eqref{eq:eliminate_spec} and \eqref{eq:generate_des}. The retroreflection efficiency is defined as $\zeta=|R|^2$. The objective function is set to
\begin{align}
    O &= \min. \{- \frac{1}{2} \sum_{m=0}^{N-1} |I_m|^2 \Re\{ Z_{{\rm L},m} \} \nonumber \\
        & \quad + \frac{N d \cos \theta_{\rm i}}{2 \eta_0}  \left| \left| E^{\rm far} \left( \theta_{\rm r} \right)\right|^2 - R^2 \left| 
E^{\rm far}_{\rm refer}(\theta_{\rm r}) \right|^2 \right| \},
\end{align} 
where the first term is used to evaluate the dissipated power (absorption), while the second term estimates the difference between the retroreflected power and the reference one.  When the incident angle is negative, only the first term is considered, to optimize the surface as an absorber that dissipates as much power as possible. While for the opposite sign of the incident angle, both of these two terms are taken into consideration. The retroreflected wave amplitude should reach the reference value, and as much power as possible should be dissipated by the loads to reduce parasitic reflections.

\begin{figure*}[!htbp]
    \centering
    \includegraphics{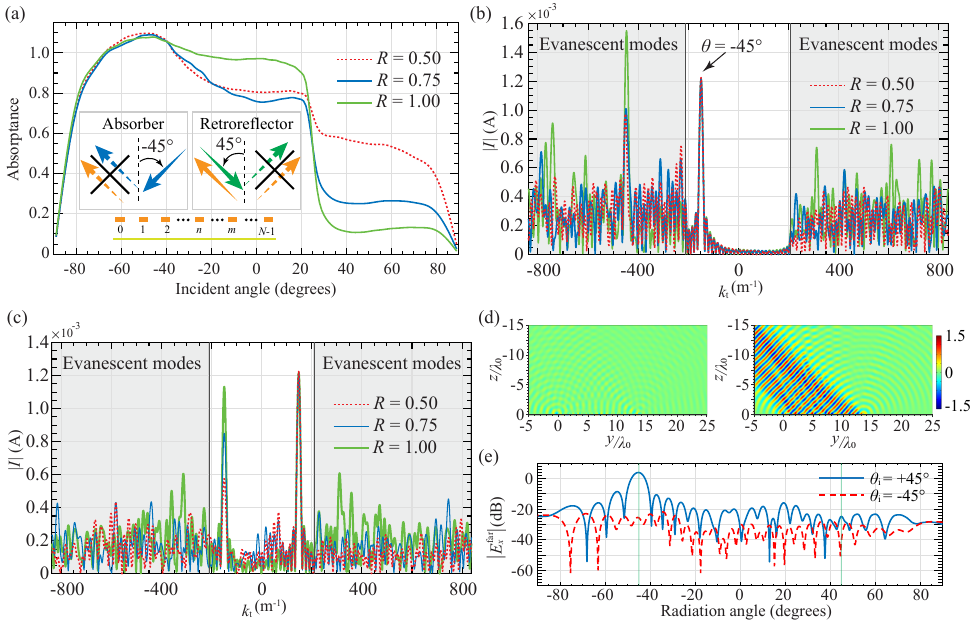}
    \caption{(a) Absorptance as a function of the incident angle for the case $M=4$. Solid red, blue, and green curves represent the reflection coefficient defined in \eqref{eq:retroreflection_coeff} with $R=0.5$, 0.75, and 1, respectively. (b) The amplitude of the induced currents as a function of the tangential component of the wavevector for $\theta_{\rm i} = -45\degree$. (c) The same for $\theta_{\rm i} = 45\degree$. The red dotted, blue solid, and green solid curves represent the target reflection coefficients defined in \eqref{eq:retroreflection_coeff}, which are 0.5, 0.75, and 1, respectively. {The shaded area represents the evanescent-wave region, where $|k_t|>k_0$}. (d) The real part of the scattered electric field $\Re\{ E^{\rm sca}_x\}$ [V/m] distribution for the case $R=1$, showing negligible reflection for illumination at $-45\degree$ (left) and high reflection for incidence at  $+45\degree$ (right). (e) Scattered electric field distribution in the far zone normalized to the maximum value of the scattered electric field in the far zone when the incident angle equals $45\degree$, with $R=1$.}    \label{fig:asymmetric_absorption}
\end{figure*}

To illustrate asymmetric superabsorption properties, we design arrays for $\theta_{\rm i}=\pm 45\degree$ and the target values of the retroreflection coefficient are equal to $R=0.5, 0.75$, and $1$. The absorptance as a function of the incident angle is shown in \figref{fig:asymmetric_absorption}(a). The asymmetric absorption spectrum is realized by controlling the evanescent wave distribution. To display the contributions of the evanescent waves, Fourier transform of the induced current is taken from the space coordinate to the tangential component of the wavevector. The amplitude of the induced currents as a function of the tangential component of the wavevector is displayed in Figs.~\ref{fig:asymmetric_absorption}(b) and (c) for incident waves at $\theta_{\rm i}=-45\degree$ and $+45\degree$, respectively. 

When waves are incident from $-45\degree$, the device works as an absorber with superabsorption for the design retroreflection coefficients. The effect of high absorption can be understood as: the induced current will generate a wave traveling toward the specular direction to eliminate the specularly reflected wave of the incident wave from the PEC ground plane.  For different retroreflection coefficients, the amplitude shows almost the same amplitude in the specular direction, as is seen in \figref{fig:asymmetric_absorption}(b), which means that the specularly reflected wave is fully eliminated. At the same time, the absorptance is obviously larger than $100\%$, which clearly shows superabsorption.

When the incident angle is $+45\degree$,  the device behaves as a retroreflector with high-level retroreflection and low absorption. The amplitude of the retroreflected wave can be controlled by the design retroreflection coefficient  \eqref{eq:retroreflection_coeff}. With the increase of the design retroreflection coefficient $R$, the amplitude of the induced current harmonic that creates a wave in that direction becomes larger, corresponding to a stronger retroreflected wave, as can be seen in \figref{fig:asymmetric_absorption}(c). The other harmonic inside the propagation part of the spectrum has a constant amplitude, as required to cancel out the specularly reflected wave from the ground plane. The asymmetry level (the ratio of the absorption of the wave coming from two opposite, mirror-symmetric directions) is tuned by the set retroreflection coefficient in \eqref{eq:retroreflection_coeff}. 

To better illustrate the angular asymmetric absorption, we calculated the scattered electric field distribution for plane-wave illuminations at $-45\degree$ and $+45\degree$ with the retroreflection coefficient of $R=1$. The results, obtained using COMSOL Multiphysics, are displayed in the left and right panels of \figref{fig:asymmetric_absorption}(d), respectively. The radiation pattern is depicted in \figref{fig:asymmetric_absorption}(e), when the incident angles equal $\pm 45\degree$. A strong retroreflection can be observed $+45\degree$ illumination, and the reflection efficiency $\zeta$ equals $100.0\%$. One can further increase this efficiency value even exceeding unity by increasing the number of $M$. The absorptance at $-45\degree$ illumination reaches $107.8\%$, which means superdirective absorption.  We can conclude that in contrast to the earlier study of asymmetric absorption in periodical arrays~\cite{wang2018extreme}, where the design angle was fixed by the geometrical period and the maximum absorptivity could not overcome 100\%, the developed global optimization of subwavelength arrays allows realization of reconfigurable asymmetric superabsorbers.

\section{Superdirective flat focusing lenses: tunable multi-focal points with adjustable intensity ratios}
\label{sec:section5}

Finally, we show possibilities to shape reflections in more general ways, in addition to discussed above manipulations of plane waves. We select an important application of focusing reflected electromagnetic waves. Designing focusing devices presents inherent challenges, such as overcoming the diffraction limit or designing lenses with high numerical apertures. Moreover, thin planar or conformal designs are preferable. These issues are essential for a wide range of applications, such as medical treatments~\cite{sharma2022development}, wireless power transfer~\cite{yu2018design}, optical imaging systems~\cite{wang2023dynamic}, and wireless communications~\cite{Li2017single}. Here, we show that it is possible to realize {reflective} reconfigurable focusing lenses, that is, focusing reflectors where the focal length, focal intensity, and the number of focal points can be dynamically controlled by tuning the loads of a subwavelength array.

We use the same platform as described above: a finite-size array of tunable impedance wires above a conducting ground plane. The design process can be formulated as follows:

Step 1: Define the desired number of focal points, the positions of the focal points, and the desired intensity ratio between the fields at different focal points.

Step 2: Define the objective function. Here, we use the following function:
\begin{equation}
    \small{
    O = \text{min.} \sum_{i=1}^{N_{\rm f}}\left \{ \left| \left| \frac{ E^{\rm sca } (\hat{y}_i,\hat{z}_i)}{E^{\rm sca } (\hat{y}_1,\hat{z}_1)}\right| - \sqrt{F_i}  \right| - \left| \frac{E^{\rm sca } (\hat{y}_1,\hat{z}_1)}{E_0} \right|  \right\}
    },
    \label{eq:obj_focus}
\end{equation}
where the number of focal points is $N_{\rm f}$, and $F_i$ is the intensity ratio between field amplitudes at the positions of focal points and the reference focal point (point number 1). The choice of the reference focal point is arbitrary. The position of the reference focal point is denoted as $(\hat{y}_1,\hat{z}_1)$. The position of the $i$th focal point is $(\hat{y}_i,\hat{z}_i)$. The first term is used to evaluate the difference between the fields at the $i$th focal point and the reference focal point. The second item is to ensure that at all focal points the input power is focused as strong as possible, i.e., to ensure high focusing efficiency. For a single focal point, i.e., when $N_{\rm f}=1$, the first term of the defined objective in \eqref{eq:obj_focus} vanishes because  $F_1=1$, and objective function optimization leads to the optimized concentration of reflected fields at the focal point. When multiple focal points are required, the objective function reaches its minimum value when the field amplitude at the reference focal point $(\hat{y}_1,\hat{z}_1)$ attains the maximum permissible value, while the fields at the other focal points simultaneously achieve the desired focal intensity ratios.

Step 3: Run optimization and evaluate the performance of the designed focusing lens. The numerical aperture (NA) is defined as $\text{NA} = n \sin \theta$~\cite{paniagua2018metalens}, where $n$ is the index of refraction of the medium in which the lens is working (1 for air), and $\theta$ is the half-angle of the maximum cone of light that can enter or exit the strip array. There are two established methods for measuring focusing efficiency (FE). The first method involves calculating the ratio between the power of the scattered wave within the focal spot (the distance between the first minima) and the total incident power~\cite{ratni2020dynamically,popov2021non}. The other method evaluates FE by calculating the ratio between the power of the scattered wave within the focal spot width (which is defined as {six} times the full width at half maximum (FWHM)) and the total incident power~\cite{ding2018graphene, arbabi2015subwavelength}. The two methods give close results, although the latter evaluation method produces a slightly higher FE. In this work, the first method is used to evaluate the FE.

In addition to this conventional characterization, we calculate the reflected power flux through the area of the focal plane (per unit length) that is equal to the area of the reflecting array. We define the reflection efficiency as the ratio of this value and the power incident
at the array surface:

\begin{equation}
    \zeta_{\rm F} = \frac{P_{\rm ref}}{P_{\rm inc}},
    \label{foc_effi}
\end{equation}
where the reflected power $P_{\rm ref}$ reads $\int_{-\frac{d}{2}}^{(N-1)d+d/2} S_z dy$, with the normal component of Poynting vector $S_z = \frac{1}{2} \Re \left\{ E_x^{\rm sca}\times \left( H_y^{\rm sca} \right)^\star \right\}$.
This value is used to judge if the focusing device exhibits superdirective properties.

\subsection{One focal point}
\begin{figure}[!htbp]
    \centering    \includegraphics{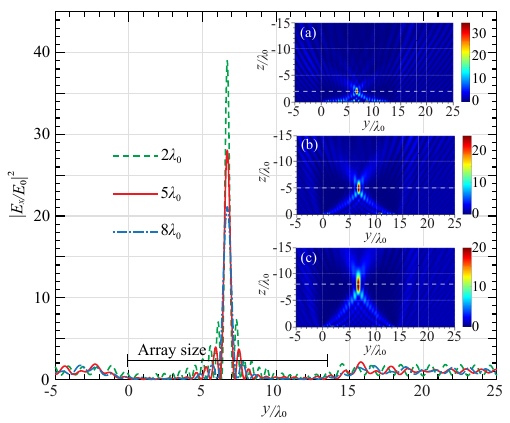}
    \caption{Intensity distribution of the reflected and scattered electric field $ \left| E_x^{\rm sca} \right|^2~[{\rm V/m}]^2$ along the focal plane. Panels (a), (b), and (c) show  the field distributions for focal points set at the symmetry axis of the strip array with the focal distances$|z_1|=2 \lambda_0$, $|z_1|=5 \lambda_0$ and $|z_1|=8 \lambda_0$, respectively.}
    \label{fig:single_point}
\end{figure}

Here, we design reflectors with three different focal lengths for normally incident waves. We set the required focal distances to $2\lambda_0, 5\lambda_0$, and $8\lambda_0$. The achieved intensity distribution of the reflected electric field distribution in the focal planes (shown by the dashed white line) for different focal lengths is depicted in \figref{fig:single_point}. The physical aperture size equals $13.5\lambda_0$. For focal length  $|z_1| = 2\lambda_0$, NA equals $0.989$, and the FWHM is $0.0109~\text{m}$ ($\approx 0.36 \lambda_0$) with FE as high as $75.7\%$. The spot size measured from the central maximum to the first minima in the diffraction pattern is $0.387\lambda_0$ (i.e., $\frac{0.383\lambda_0}{\rm NA}$), which is smaller than the size of the Airy spot, $\frac{0.61\lambda_0}{\rm NA}$ (the Rayleigh
criterion)~\cite{rayleigh1896xv, huang2018planar,popov2021non}. This property demonstrates subdiffraction focusing. Meanwhile, the spot size value is very close to the superoscillation criterion value, which is known as the minimum spot size of a subdiffraction focal spot  ($\frac{0.38\lambda_0}{\rm NA}$)~\cite{huang2014optimization,huang2018planar}. For the focal length $|z_1| = 5 \lambda_0$, NA reads $0.938$, and the spot size is $\frac{0.472\lambda_0}{\rm NA}$, which is also smaller than the size of the Airy spot. The FWHM is about $0.0140~\text{m}$ ($\approx 0.47 \lambda_0$), and the FE is as high as $83.7\%$.  For the focal length $|z_1|=8\lambda_0$,  NA reads $0.860$, and the spot size is $\frac{0.576\lambda_0}{\rm NA}$, which is still smaller than the size of the Airy spot. The FWHM is about $0.0183~\text{m}$ ($\approx 0.61 \lambda_0$), and the focusing efficiency is also as high as $89.5\%$. The FWHM becomes wider as the focal length increases, and the sidelobes become smaller. As a result, the FE becomes higher. Interestingly, the side-lobe level is rather uniformly minimized in the focal plane over the same area as the area of the array.

When the focal length equal $2\lambda_0$, $5\lambda_0$, and $8\lambda_0$, the corresponding reflection efficiencies  $\zeta_{\rm F}$ \eqref{foc_effi} are $101.9\%$, $106.2\%$, and $106.3\%$, respectively. This result shows that in all these cases the device is superdirective: the reflected flux is higher than the incident flux.
A comparison of the performance between this work and some of the known comparable designs is shown in \tabref{tab:comparison1}. {Note that the comparison with previous works in TABLE I is based on theoretical performance, as the proposed design assumes ideal reconfigurable loads without ohmic loss. }

\begin{table}[ht]
    \caption{Performance comparison of some previous works with the presented reflectarray realization}
    \centering
    \begin{tabular}[c]{ccccc}
    \toprule
    References     & \thead{Frequency \\(GHz)}  & \thead{Focal length \\ ($\lambda_0$)} & \thead{FWHM \\ ($\lambda_0$)}  & FE (\%) \\
    \midrule
    \multirow{3}{*}{Y.-Q. Liu et al.~\cite{liu2020high}} & 7.5 & 3.43 & 0.9 & $\approx 39$  \\
    & 9         & 1.89         &0.57  & $\approx 48$ \\
    & 9.5        & 2.41         &0.54  &  $\approx 34$ \\
\cline{2-5}
\multirow{3}{*}{V. Popov et al.~\cite{popov2021non}} & \multirow{2}{*}{9.5} & 1 & 0.43& 21\\
& & 3 & 0.6 & 30 \\
& 10 &  $3$& 0.6 & 57 \\
\cline{2-5}
\multirow{3}{*}{B. Ratni et al.~\cite{ratni2020dynamically}} & \multirow{3}{*}{9}& 1.5 & 0.53 &42\\
& & 2.25 & 0.6 & 40\\
& & 3 & 0.69 & 33\\
\cline{2-5}
    \multirow{3}{*}{This work} & \multirow{3}{*}{10} & $2$  & 0.36 & 75.7 \\
                              &                                      & $5$ & 0.47 & 83.7  \\
                              &                                      & $8$ & 0.61 & 89.5  \\
    \bottomrule
    \end{tabular}
    \label{tab:comparison1}
\end{table}

\begin{figure}[ht]
    \centering
    \includegraphics{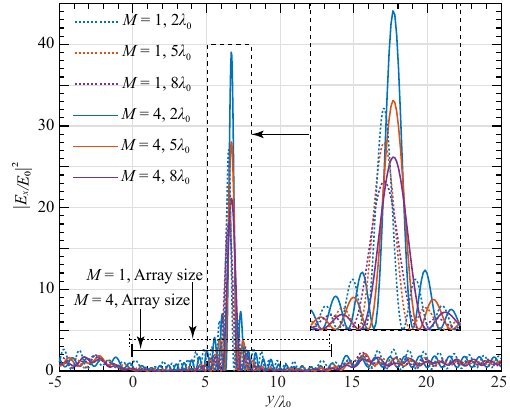}
    \caption{Intensity of the scattered electric field $ \left| E_x^{\rm sca} \right|^2~[{\rm V/m}]^2$ distribution over the  focal plane. The blue, red, and purple curves correspond to the strip arrays with the focal length $|z_1|=2 \lambda_0$, $|z_1|=5 \lambda_0$, and $|z_1|=8 \lambda_0$, respectively. A slight shift in the maxima position is attributed to the change of the array symmetry axis, which arises from alterations in the number of strips $N$.}
    \label{fig:single_point_M=1_M=4}
\end{figure}

It is important to compare the achieved performance with that of conventional reflectarrays formed by $\lambda_0/2$-spaced elements. To do that, we perform the same design optimizations for one strip in each $\lambda_0/2$ section ($M=1$) and compare it with the above presented designs with four elements per $\lambda_0/2$ ($M=4$). The intensity distributions  of the scattered electric fields along the focal plane are depicted in~\figref{fig:single_point_M=1_M=4}. We see that as the number  $M$ increases, the field intensity at the focal point is significantly enhanced, also the FE and $\zeta_{\rm F}$ can be significantly improved. However, the spot size and the FWHM show no obvious changes. 
The realized parameters are compared in~\tabref{tab:comparison2}. We see that superdirective focusing lens cannot be realized when $M=1$, because the reflection efficiency is in all cases smaller than 100\%.

\begin{table}[ht]
    \caption{Effect of the array period (number of strips $M$ per $\lambda_0/2$) on the achievable focusing performance}
    \centering
    \begin{tabular}[c]{cccccc}
    \toprule
        \thead{Focal length \\ ($\lambda_0$)}& 
        $M$ & \thead{FWHM \\ ($\lambda_0$)} & \thead{Spot size \\ ($\lambda_0$)}  & FE ($\%$) & $\zeta_{\rm F} ($\%$)$\\
        \midrule
        \multirow{2}{*}{$2$}
        & 1 & 0.36 & 0.375 & 50.7 & 95.1\\
        & 4 & 0.36 & 0.387 & 75.7 & 101.9\\
        \cline{2-6}
        \multirow{2}{*}{$5$}
        & 1 & 0.49 & 0.530 & 72.9 & 96.6\\
        & 4 & 0.47 & 0.503 & 83.7 &106.2\\
        \cline{2-6}
        \multirow{2}{*}{$8$}
        & 1 & 0.64 & 0.700 & 80.7 & 96.0\\
        & 4 & 0.61 & 0.670 & 89.5 & 106.3\\
        \bottomrule
    \end{tabular}
    \label{tab:comparison2}
\end{table}

\subsection{Different incident angles}
\begin{figure}[h]
    \centering
    \includegraphics{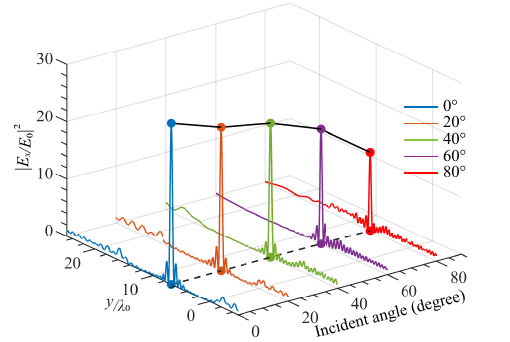}
    \caption{Focal-plane intensity distributions of the scattered electric field for different incident angles. The focal length is fixed to $\left| z_1 \right| = 5\lambda_0$.}
    \label{fig:different_angles}
\end{figure}

For conventional focusing lenses, the focal point position shows some shift when the angle of incidence changes. Here we show that the focal point can be kept at the same position for different incident angles by properly adjusting the impedance loads for each incident angle. Upon such tuning, the only difference of performance for different illumination angles is the intensity value at the focal point which decreases with increasing incident angle because of the decreasing incident power at tilted illumination angles. These results are presented in \figref{fig:different_angles}.

\subsection{Two focal points}

\begin{figure}[!htbp]
    \centering    \includegraphics{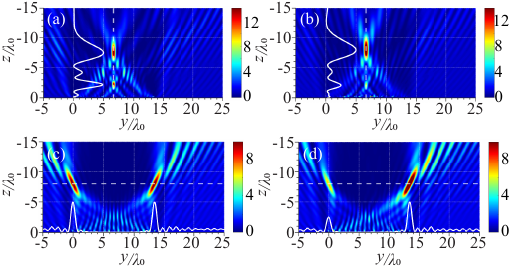}
    \caption{(a)--(b) Intensity distribution of the scattered electric field $ \left| E_x^{\rm sca} \right|^2~[{\rm V/m}]^2$ distribution, where the two focal points are on the central axis of the strip array at the focal lengths $2 \lambda_0$ and $8 \lambda_0$.  (c)--(d). The two focal points are at $(0, -5 \lambda_0)$ and $(N-1)d, -5\lambda_0)$. The solid white lines represent the intensity of the scattered electric field normalized to its maximal value, that is, $\small{ \left| E^{\rm sca} \right|^2 \over \text{max.}\{\left| E^{\rm sca} \right|^2\} }$.}
    \label{fig:two_focal_points}
\end{figure}

Next, we present examples that show possibilities of realizing multi-point focusing. 
In the first case, we design reflectors with two focal points along the symmetry axis of the array. The array dimensions and position are the same as in the above examples. We set the positions of the two focal distances to $2\lambda_0$ and $8\lambda_0$ and the desired ratios of the field intensities at the focal points to  1:1 or 1:2. 
The results are shown in~\figref{fig:two_focal_points}(a) and (b). The normalized intensity curves (white solid lines) show that the desired intensity ratio is indeed realized.
 
Next, we design arrays with two off-centered positions of the two focal points. The desired two focal points are set at $(0, -5 \lambda_0)$ and $((N-1)d, -5\lambda_0)$. The results are depicted in~\figref{fig:two_focal_points}(c) and (d) for the intensity ratios  1:1 and 1:2, respectively.

\section{Conclusion}
To conclude, we have demonstrated a general method for controlling the wavefronts of electromagnetic waves, overcoming the limit of 100\% efficiency of conventional devices. This promising approach allows for the design of reconfigurable and multifunctional devices,  all integrated within a single platform. The presented examples include beam splitters, focusing lenses, and absorbers, and we have shown that all of them are capable of reflection or absorption of more power than is incident on their surfaces. To determine the necessary load impedances that align with the desired functionalities, we employ the global optimization method known as MPGA. Our nonlocal design strategy exhibits dynamic adaptability, significantly enhancing the ability to engineer electromagnetic waves. 

Comparisons with known designs of similar devices demonstrate superior performance in most demanding settings. The multi-angle reflectors show outstanding freedom in manipulating the power flow toward arbitrary directions with arbitrary power ratios between different reflection angles with superdirective efficiency. The designed focusing lenses show numerical apertures and high focusing efficiencies even for extremely small focal distances.  When the angle of incidence changes, the array loads can be adjusted to keep the focal point at the same position. Moreover, this approach can be extended to designing focusing lenses with several focal points and relative intensities for different focal points can be controlled at will. For absorber design, we show superdirective absorption in an extremely wide angular sector, between approximately $\pm 85\degree$. This functionality can be realized by setting multiple incident angles for optimized symmetric absorption. Furthermore, we show that for angle-asymmetric absorption, the evanescent waves can be controlled for different incident angles, also reaching superabsorptive performance for the desired angular sector of absorption. 

We expect that our explorations can be used in the next generation of communication systems, allowing creation of extremely effective and fully reconfigurable metausurfaces using simple universal platforms. Here, our conceptual study was made for arrays of impedance-loaded strips or wires, allowing scanning only in one plane. In related recent works \cite{Sravan1,Sravan_arxiv} it was shown that similar optimization methods can be used for more practical arrays of patch antennas loaded by bulk tunable loads. In future works, we plan to generalize the method for two-dimensional arrays, allowing complete control over reflection into any direction and two-dimensional scans into any angle without parasitic scattering and pattern deterioration.   As research in this field progresses, we can anticipate a plethora of applications across diverse sectors, from communications to energy harvesting and beyond.

\section*{Acknowledgments}
The authors thank Prof. Constantin Simovski, Dr. Grigorii Ptitcyn, and Dr. Francisco Cuesta for helpful discussions.


{\appendices
\section{Derivation of the Required Induced Current}

Here we show how to find the values of the induced currents that create the adopted reference reflected field for ideal but not superdirective anomalous reflectors. In this case, the desired reflected beam is created by a uniform-amplitude current with a linearly varying phase. 

In practice, thin conducting strips are modeled as equivalent round wires with the effective radius $r_{\rm eff} = w/4$~\cite{tretyakov2003analytical}. For a single current line $\mathbf{J}= I \delta(y_0, z_0)\hat{x}$ positioned at $(y_0,z_0)$ and parallel to $x$-axis, the scattered electric field at  point $(y,z)$ can be calculated as~\cite{tretyakov2003analytical}
\begin{equation}
    \mathbf{E}_x = -\frac{k_0 \eta_0 }{4} I H_0^{(2)}(k_0 \sqrt{(y-y_0)^2 + (z-z_0)^2}) \hat{x},   
\end{equation}
where $k_0 = \omega \sqrt{\mu_0 \epsilon_0}$ is the wavenumber in free space, $\eta_0 = \sqrt{\mu_0 / \epsilon_0}$ represents the wave impedance in free space, and $H_0^{(2)}(\cdot)$  is the zeroth-order Hankel function of the second kind.

In order to eliminate the specular reflection and generate the desired reflected plane wave, two sets of induced currents are required. Let us denote one set of currents that eliminates specular reflection as $\mathbf{J}_\alpha(y, z) = \sum_{m=0}^{N-1}{I_{\alpha} \over d} e^{-j k_0 \sin \theta_{\rm i} y_m}\delta(y - y_m, z + h) \hat{x}$, and the other set of currents that creates the desired anomalously reflected wave as $\mathbf{J}_\beta(y, z) = \sum_{n=1}^{N_{\rm c}}\mathbf{J}_{\beta_n}(y, z) = \sum_{n=1}^{N_{\rm c}} \sum_{m=0}^{N-1} {I_{\beta_n} \over d}e^{-j k_0 \sin \theta_{{\rm r}_n} y_m}  \delta(y - y_m, z + h) \hat{x}$. Because both current components generate plane waves (in the limit of an infinite array, the complex amplitudes of the strip currents $I_{\alpha,\beta_n}$ ($n=\{ 1,2,\dots, N_{\rm c}\}$) are the same for all strips.

\subsection{Elimination of the specularly reflected wave}
To eliminate specular reflection from the ground plane, the sum of the surface-averaged value (the fundamental Floquet harmonic) of the scattered field generated by the induced strip currents and the incident field reflected from the ground should equal zero. We write this condition at $z=-h$, just above the array:  $\mathbf{E}_{\alpha} + \mathbf{E}_{\rm ref} (y,-h) = 0$. 
The electric field generated by the strips can be found using \cite[Eq.~4.35]{tretyakov2003analytical}, which gives 
$
  -\frac{\mathbf{J}_\alpha \eta_0}{2 \cos \theta_{\rm i}}  $. 
The total radiated field is the sum of that electric field generated by the strips and the electric field reflected from the ground:
\begin{equation}
    \mathbf{E}_\alpha = -\frac{\mathbf{J}_\alpha \eta_0}{2 \cos \theta_{\rm i}}  \left( 1 - e^{- j k_0 \cos \theta_{\rm i} 2h} \right).
\end{equation}
Substituting $\mathbf{E}_{\rm ref} (y,-h)$ from \eqref{eq:reflected} and writing $\mathbf{J}_\alpha={I_\alpha\over d}\hat x$, we find 
\begin{equation}
    I_\alpha = j \frac{E_0 d \cos \theta_{\rm i} }{\eta_0 \sin (k_0 \cos \theta_{\rm i} h)}.
    \label{eq:I_alpha}
\end{equation}

\subsection{Generation of the desired reflected wave}
The averaged scattered electric field in angle $n$ generated by the current component $I_{\beta_n}$ reads, similarly, 
\begin{equation}
    \mathbf{E}_{\beta_n} = -\frac{\mathbf{J}_{\beta_n} \eta_0}{2 \cos \theta_{{\rm r}_n}} (1 - e^{- j k_0 \cos \theta_{{\rm r}_n} 2h}).  
\label{eq:electric_field_beta}
\end{equation}
From this relation, we see that the distance between the plane of strips and the ground plane should satisfy  $h \ne q \frac{\lambda_0}{2\cos \theta_{{\rm r}_n}}$, where $q$ is a positive natural number. Otherwise, the sum of the primary field of these currents and the reflection from the ground would equal zero, $\mathbf{E}_{\beta_n} = 0$, and we could never generate the desired plane wave in angle $n$.

To find the required amplitude of current $\mathbf{J}_{\beta_n}$, we use the relationship between the power intensity between angle $n$ and the incident power intensity, equating the normal components of the incident plane wave and the anomalously reflected plane wave. First, we calculate the tangential component of the magnetic field in angle $n$ that corresponds to the plane wave with the electric field \eqref{eq:electric_field_beta}. This gives  
\begin{equation}
    \mathbf{H}_{\beta_n} \cdot \hat y= \frac{E_{\beta_n}} {\eta_0 }\cos \theta_{{\rm r}_n}.
    \label{eq:magnetic_field}
\end{equation}
The corresponding normal component of the Poynting vector reads $\frac{|E_{\beta_n}|^2}{2 \eta_0}\cos\theta_{{\rm r}_n}$. 
The normal component of the incident-wave Poynting vector reads $\frac{|E_{\rm inc}|^2}{2 \eta_0}\cos\theta_{\rm i}$. According to the power intensity ratio between angle $n$ and the incident power intensity shown in \eqref{eq:power_ratio}, we find the required current amplitude 
\begin{equation}
    \left| {I_{\beta_n}} \right| = |J_{\beta_n} |d = \sqrt{\rho_n}  \left| \frac{E_0 d \sqrt{\cos \theta_{\rm i} \cos \theta_{{\rm r}_n}}} {\eta_0 \sin (k_0 h \cos \theta_{{\rm r}_n})} \right|.
    \label{eq:I_beta}
\end{equation}

\section{Configuration of the COMSOL Multiphysics simulation for finite-sized strip array}
The background field $\mathbf{E_{\rm b}}=\mathbf{E_{\rm inc}} = e^{-j k_0 \sin \theta_{\rm i} y - j k_0 \cos \theta_{\rm i} z} \hat{x}$ is used to excite the system. We calculate the corresponding loads using \eqref{eq:matrix_Ohm} and insert these values into commercial software COMSOL Multiphysics to calculate the scattered field. The configuration of the COMSOL simulation setting is displayed in \figref{fig:COMSOL_configuration}.
\begin{figure}[!htbp]
    \centering
    \includegraphics{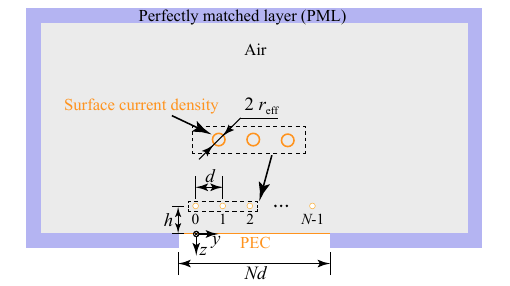}
    \caption{Schematic of the 2D Comsol Multiphysics simulation configuration. The blue area is the perfectly matched layer (PML). The first strip is placed at the positions $y=0$ and $z=-h$. The surface current density on the wires is defined in terms of an impedance boundary condition (see {orange} hollow circle), as $\mathbf{J} = E_x/\left(  Z_{{\rm L},n} / (2 \pi r_{\rm eff}) \right) \hat{x}$. {Here, the insertion period $l$ is not specified in the simulation, as $l$ defines the load impedance per unit length  $Z_{{\rm L},n}$, which is measured in   $\rm{\Omega/m}$.}}
    \label{fig:COMSOL_configuration}
\end{figure}

%
\bibliographystyle{IEEEtran}
\bibliography{IEEEabrv,refs}

\end{document}